\documentclass[superscriptaddress,groupedaddress,nofootnoteinbib,12pt]{article}
\usepackage{color}
\usepackage{graphicx}
\usepackage{dcolumn}
\usepackage{bm}
\usepackage{amssymb}
\usepackage{amsmath}
\usepackage{sectsty}
\usepackage{colortbl}
\usepackage{latexsym}
\usepackage{mathrsfs}
\usepackage{float}
\usepackage{ifthen}
\usepackage{caption,subfig}
\usepackage{enumerate}
\usepackage{url}
\usepackage{caption,subfig}	
\usepackage{jcappub}

\def\be{\begin{equation}}
\def\ee{\end{equation}}
\def\ba{\begin{eqnarray}}
\def\ea{\end{eqnarray}}
\def\beq{\begin{eqnarray}}
\def\eeq{\end{eqnarray}}

\def\mpl{M_{\rm Pl}}

\def\d{\mathrm{d}}

\def\K{{\cal K}}

\def\L*{{\cal L}_*}
\def\L{\mathcal{L}}
\def\({\left(}
\def\){\right)}

\def\<{\langle}
\def\>{\rangle}

\def\cs2{c_{s}^{2}}

\def\be{\begin{equation}}
\def\ee{\end{equation}}
\def\ba{\begin{eqnarray}}
\def\ea{\end{eqnarray}}
\def\beq{\begin{eqnarray}}
\def\eeq{\end{eqnarray}}

\def\mpl{M_{\rm Pl}}

\def\d{\mathrm{d}}

\def\K{{\cal K}}

\def\L*{{\cal L}_*}
\def\L{\mathcal{L}}
\def\({\left(}
\def\){\right)}

\def\<{\langle}
\def\>{\rangle}

 \def\be   {\begin{equation}}   \def\ee   {\end{equation}}

 \def\ba  {\begin{eqnarray}}   \def\ea  {\end{eqnarray}}

\renewcommand{\thesubsection}{\arabic{section}.\arabic{subsection}}

\begin{document}
\hspace{5.2in} \mbox{NORDITA-2015-12}\\\vspace{-1.03cm} 

\title{Revisiting perturbations in extended quasidilaton massive gravity}

\author{Lavinia Heisenberg$^{a,b}$}
\affiliation{$^{a}$Nordita, KTH Royal Institute of Technology and Stockholm University, \\
Roslagstullsbacken 23, 10691 Stockholm, Sweden}
\affiliation{$^{b}$Department of Physics \& The Oskar Klein Centre, \\
AlbaNova University Centre, Roslagstullsbacken 21, 10691 Stockholm, Sweden}

\emailAdd{Lavinia.Heisenberg@kth.se}

\abstract{In this work we study the theory of extended quasidilaton massive gravity together with the presence of matter fields. After discussing the homogeneous and isotropic fully dynamical background equations, which governs the exact expansion history of the universe, we consider small cosmological perturbations around these general FLRW solutions. The stability of tensor, vector and scalar perturbations on top of these general background solutions give rise to slightly different constraints on the parameters of the theory than those obtained in the approximative assumption of the late-time asymptotic form of the expansion history, which does not correspond to our current epoch. This opens up the possibility of stable FLRW solutions to be compared with current data on cosmic expansion with the restricted parameter space based on theoretical ground.}

\maketitle

\section{Introduction}

It has been a long time since we know that the universe is expanding, which means that if we look at two distant galaxies, the separation between them increases in time. The rate at which the galaxies are getting apart is determined by the content of the universe and the current consensus is that ordinary matter (baryons, electrons ... etc) only represents about five percent of the content of the universe. A larger fraction, about 27\%, corresponds to what is known as dark matter. This is a component that behaves like a dust fluid but does not interact (or at least very weakly) with photons. Its existence is however inferred by its gravitational effects, ranging from galactic scales to the universe horizon. It is also crucial for the formation of the structures that we observe today. Finally, maybe the most intriguing component is what is usually called dark energy, which amounts to about 70\% of the universe content and it is responsible for the accelerated expansion of the universe. This accelerated expansion of the universe was first inferred by measurements of type Ia supernovae. These supernovae are standarizable candles, that can be used to measure cosmological distances. One of the pillars of the standard model of cosmology is the Cosmological Principle, which states that the universe is homogeneous and isotropic on large scales. These symmetries impose that the metric of the universe must be of the Friedmann-Lema\^itre-Robertson-Walker (FLRW) form, where the scale factor $a(t)$ determines physical distances. Thus, by measuring the distance to supernovae at different redshifts, we can have a measure of the expansion rate of the universe. In General Relativity, from Einstein equations one can obtain the acceleration equation (second derivative of the scale factor). It is immediately clear that in order to have an accelerated expansion, the universe must be dominated by a fluid satisfying the condition $\ddot{a}>0$, or in other words the pressure of the fluid must fulfil the condition $P<-\frac{1}{3}\rho$. Thus, this condition is not satisfied by the usual fluids like dust or radiation.\\

The simplest way to achieve acceleration is the introduction of a cosmological constant $\Lambda$. This gives a contribution in the form of a perfect fluid with $P_\Lambda=-\rho_\Lambda$, and, therefore, provides acceleration. This has become the concordance model or standard model of cosmology, called $\Lambda$CDM. It is able to fit most cosmological observations with a high accuracy. We also have measurements of large scale structures where we measure the distribution of galaxies in the universe, and $\Lambda$CDM also gives good fits to current data. This type of measuremets will be the next generation of cosmological observations. Although the cosmological constant is in good agreement with current observations, it has some problems from a theoretical viewpoint. First of all, if $\Lambda$ is a true constant of nature belonging to the gravitational sector, then the gravitational action has two dimensionful constants which differ by many orders of magnitude and this makes it unnatural. On the other hand, the cosmological constant will also receive radiative corrections from quantum fields, and those corrections are again many orders of magnitude larger than the observed value, so we need to finely tune the bare value to be in agreement with observations. The unnatural problems of the cosmological constant and its instability against quantum corrections motivated searching for alternatives to a cosmological constant in order to explain the cosmic acceleration. Such alternatives can be broadly divided into two categories. On one hand, models that rely on the introduction of scalar fields, vector fields, new exotic fluids, etc as the cause of the accelerated expansion. On the other hand, there are also models that resort to modifying GR at large scales, arguing that the accelerated expansion might be due to a break down of Einstein's General Relativity on cosmological scales.\\

One interesting way of modifying gravity in the infra-red (IR) is massive gravity. The simplest linear massive gravity theory is the pioneering model proposed by Fierz and Pauli \cite{Fierz:1939ix}. The nonlinear covariant completion of the Fierz-Pauli theory with the correct degrees of freedom was formulated by de Rham, Gabadadze and Tolley (dRGT) \cite{deRham:2010ik,deRham:2010kj}. The potential is tuned in a way that guarantees the absence of the Boulware--Deser (BD) ghost \cite{Boulware:1973my}. This relative tuning is radiatively stable in the decoupling limit of the theory due to  the antisymmetric structure of the potential interactions \cite{deRham:2012ew} (Galileon interactions share this nice property as well \cite{Luty:2003vm,Nicolis:2004qq,Hinterbichler:2010xn,deRham:2012ew,Heisenberg:2014raa}). The radiative stability does not survive beyond the decoupling limit, however the introduced BD ghost is harmless since its mass is beyond the cut-off scale of the theory \cite{deRham:2013qqa}. Additionally, the quantum corrections arising from matter loops might dictate the form of allowed classical interactions with matter fields \cite{deRham:2014naa,deRham:2014fha,Heisenberg:2014rka}. The phenomenology of the dRGT theory is reach. Its potential impact on cosmology is a very interesting question \cite{deRham:2010tw,PhysRevD.84.124046,PhysRevLett.109.171101,deRham:2011by,Chamseddine:2011bu,Koyama:2011xz,Koyama:2011wx,Gumrukcuoglu:2011zh,Gratia:2012wt,Vakili:2012tm,Kobayashi:2012fz,Fasiello:2012rw,Volkov:2012zb,Tasinato:2012ze,Wyman:2012iw,Gratia:2013gka,DeFelice:2013bxa,Fasiello:2013woa,Heisenberg:2014kea,Comelli:2013tja,Motloch:2014nwa}. The construction of a realistic and stable cosmology has witnessed many promising attempts, even though most of them were doomed to a failure very quickly. The first failure was the construction of flat FLRW solution \cite{PhysRevD.84.124046}. Similarly, even if it was possible to find self-accelerating open FLRW solutions, these were suffering from a nonlinear ghost instability \cite{PhysRevLett.109.171101}. Attempts to promote the reference metric  to de Sitter or FLRW did not help since they face the unavoidable appearance of Higuchi ghost \cite{Fasiello:2012rw}. Possible ways out of these difficulties consist on breaking the FLRW symmetries \cite{PhysRevD.84.124046,Gumrukcuoglu:2012aa,Koyama:2011xz,Koyama:2011yg,Chamseddine:2011bu,Gratia:2012wt,Kobayashi:2012fz,Volkov:2012cf} or adding additional new degrees of freedom \cite{Huang:2012pe,PhysRevD.87.064037,DeFelice:2013tsa}. One interesting way of extending the original massive gravity theory is the inclusion of the kinetic term for the reference metric \cite{Hassan:2011zd}. Another promising way is the coupling matter fields through an effective composite metric \cite{deRham:2014naa,Gumrukcuoglu:2014xba,Solomon:2014iwa}.\\

In this work we will go along the line of the presence of additional degrees of freedom. To be precise, we will focus on the extension 'quasidilaton' in form of an additional scalar field with a specific coupling to the massive graviton \cite{PhysRevD.87.064037}. The Lagrangian is constructed in such a way that it is invariant under the quasidilaton global symmetry. Even if the original version of the quasidilaton fails to provide stable self-accelerating solutions, there is a promising extension of it proposed by de Felice and Mukohyama \cite{DeFelice:2013tsa} through the addition of a new coupling constant. Interestingly enough, the background dynamics is independent of this new coupling, however for the stability of the perturbations it does play a crucial role. In the standard formulation of massive gravity, phenomenological consistency requires a small graviton mass, of the same order as the Hubble expansion rate today $m\sim H_0$. As a consequence the effective mass of tensor perturbations would be also constrained to be of the same order. In the quasidilaton scenario the effective mass of gravitational waves has an additional dependence on the quasidilaton parameter and hence it can be made much larger than the Hubble parameter. This might provide an explanation for the large-angle suppression of power in the microwave background to be explored \cite{Kahniashvili:2014wua}.\\

The scalar perturbations of the extended quasidilaton with matter fields were studied in \cite{Motohashi:2014una}. In their analysis, they assumed the late-time asymptotic form of the expansion history to evaluate the stability of scalar perturbations in the presence of matter fields. However, the late-time asymptotic solution does not correctly describe our current epoch, which is just in the transition between matter domination and accelerating expansion. In this work, we consider the full dynamical equations of motion and study the tensor, vector and scalar perturbations on top of these general background equations. By doing so, we obtain new stability constraints which highly depend on the evolution of background quantities, which were set to zero or to some constant values in the asymptotic solution.
We first review the extended quasidilaton generalization of dRGT theory and setup our Lagrangian and conventions in Section \ref{sec:quasidilatondRGT}. After calculating the fully dynamical background equations of motion on FLRW space-time in Section \ref{sec:background_evolution}, we pay special attention to the stability of the tensor, vector and scalar perturbations in the presence of the matter fields in Section \ref{sec:perturbations}, where we obtain new highly non-trivial stability constraints on the parameters of the theory. Finally, we summarise our results in Section \ref{sec:conclusion}.\\

Throughout the paper, we will work with the metric signature convention $(-,+,+,+)$ and define the reduced Planck mass as $M_{\rm Pl}=1/\sqrt{8 \pi G}$. Furthermore, the traces of some rank-2 tensors in Section \ref{sec:quasidilatondRGT} are denoted by $[...]$, for example ${\cal K}^{\mu}_{~\mu}=[{\cal K}]$,~ ${\cal K}^{\mu}_{~\nu}{\cal K}^{\nu}_{~\mu}=[{\cal K}^2]=({\cal K}_{\mu\nu})^2$,~ ${\cal K}^{\mu}_{~\nu}{\cal K}^{\nu}_{~\rho}{\cal K}^{\rho}_{~\mu}=[{\cal K}^3]=({\cal K}_{\mu\nu})^3$~etc. 


\section{Extended quasidilaton dRGT massive gravity}
\label{sec:quasidilatondRGT}
In this section we will first review the interactions in the theory of extended quasidilaton massive gravity and setup the framework in which we will perform our analysis of cosmological perturbations. Our starting point is the action for extended quasidilaton massive gravity and the matter action where the ordinary matter fields still couple minimally to the physical metric $g$ and an additional scalar field $\sigma$, the quasidilaton field, comes in the potential interactions in a very specific way such that the absence of the BD ghost is maintained \cite{DeFelice:2013tsa} 
\begin{equation}\label{action_MG_effcoupl}
\mathcal{S} = \int \mathrm{d}^4x \big[ \frac{\mpl^2}{2} \sqrt{-g}\left(R[g]-2\Lambda-\frac{\omega}{M_{\rm Pl}^2}\partial_\mu\sigma\partial^\mu\sigma+2m^2(\mathcal{U}_2+\alpha_3 \mathcal{U}_3+\alpha_4 \mathcal{U}_4) \right)+\mathcal{L}_{\rm matter}\big]\,,
\end{equation}
where the potential interactions are given by \cite{deRham:2010ik,deRham:2010kj}
\begin{eqnarray}
\mathcal{U}_2[\mathcal{K}] &=& \frac{1}{2}\left( [\K]^2-[\K^2]\right), \nonumber\\
\mathcal{U}_3[\mathcal{K}] &=& \frac{1}{6}\left( [\K]^3-3[\K][\K^2]+2[\K^3]\right),  \nonumber\\
\mathcal{U}_3[\mathcal{K}] &=& \frac{1}{24}\left([\K]^4-6[\K]^2[\K^2]+3[\K^2]^2+8[\K][\K^3]-6[\K^4]\right),
\end{eqnarray}
The tensor $\K$ is the building block of dRGT theory \cite{deRham:2010kj} with a highly non-trivial structure in form of a square root. In the extended quasidilaton massive gravity this tensor is promoted to
\begin{equation}
\K^\mu _\nu[g,f] =\delta^\mu_\nu -e^{\sigma/M_{\rm Pl}} \left(\sqrt{g^{-1}f}\right)^\mu_\nu \,,
\end{equation}
The presence of the potential interactions in \ref{action_MG_effcoupl} breaks the diffeomorphism invariance completely. Naturally, one can restore it introducing four St\"uckelberg fields $\mathcal{S}^a$ which promote the Minkowski reference metric to the space-time tensor
\begin{equation}\label{Stueckelbergfields}
f_{\mu\nu} \to \eta_{ab}\partial_\mu \mathcal{S}^a \partial_\nu\mathcal{S}^b-\frac{\alpha_\sigma}{M_{\rm Pl}^2m^2}e^{-2\sigma/M_{\rm Pl}}\partial_\mu\sigma\partial_\nu\sigma\,.
\end{equation}
As you can see the standard form of the reference metric $f_{\mu\nu} \to \eta_{ab}\partial_\mu \mathcal{S}^a \partial_\nu\mathcal{S}^b$ is promoted to \ref{Stueckelbergfields}. Note that this disformal transformation to the fiducial metric introduces a new parameter. The building block tensor $\K$ is invariant under the global symmetry $\sigma \to \sigma + \sigma_0$ and $\mathcal{S}^a \to e^{-\sigma_0/\mpl}\mathcal{S}^a$. The parameter $\alpha_\sigma$ in the disformal factor was introduced in order to render the self-accelerating late-time asymptotic solutions stable \cite{DeFelice:2013tsa}.
For sake of simplicity we will consider a matter field $\phi$ that only couples to the dynamical metric $g$ with a generic kinetic term of the form 
\begin{equation}
\mathcal{L}_{\rm matter} =\sqrt{-g}\,P(X_\phi)\,,
\label{eq:chiaction}
\end{equation}
where $X_\phi \equiv -g^{\mu\nu}\partial_{\mu}\phi\partial_{\nu}\phi$ and the energy density $ \rho_M$, the pressure $P_M$ and the sound speed $ c_M$ of the matter field can be written as
\begin{equation}
 \rho_M \equiv 2P(X_\phi)'X_\phi - P(X_\phi), \qquad  P_M \equiv P(X_\phi), \qquad
 c_M^2 \equiv \frac{P(X_\phi)'}{2P(X_\phi)''X_\phi+P(X_\phi)'}\,,
\end{equation}  
where prime corresponds to derivative with respect to the argument.
In our setup in the Lagrangian \ref{action_MG_effcoupl}, we neglected the tadpole contribution $\mathcal{U}_1$ to the mass term by setting $\alpha_1=0$, while absorbing the $\mathcal{U}_0$ term into the bare cosmological constant $\Lambda$.


\section{Background evolution}\label{sec:background_evolution}

The original formulation of massive gravity does not accommodate a simple realisation of viable cosmology due to a no-go result for flat FLRW solutions \cite{PhysRevD.84.124046}. This arises from the constraint equation imposed by the St\"uckelberg field equation of motion, which demands the scale factor to be constant. In order to avoid this no-go result, extensions to the original formulation were considered in the literature. Promising attempts are quasidilaton and extended quasidilaton formulation of the dRGT theory where an additional scalar field $\sigma$ was added. We would like to study perturbations on top of FLRW backgrounds of the latter case in the presence of matter fields coupled to the dynamical metric. Thus, we will start with the Ansatz for the dynamical metric as the homogeneous and isotropic flat FLRW metric
\begin{equation}
ds_g^2=-N^2 dt^2 +a^2 \delta_{ij} dx^idx^j\,,
\end{equation}
while the non-dynamical metric is the pull-back of the Minkowski metric in the St\"uckelberg field space to the physical space-time, which we parametrize as
\begin{eqnarray}
ds_f^2= f_{\mu\nu}dx^\mu dx^\nu &=& n^2 dt^2 +  \delta_{ij} dx^idx^j \nonumber \\
&=& \left(-\dot{f}^2-\frac{e^{-2\bar{\sigma}/M_{\rm Pl}}\alpha_\sigma \dot{\bar{\sigma}}^2}{m^2M_{\rm Pl}^2}\right)dt^2 +  \delta_{ij} dx^idx^j\,.
\label{eq:reference}
\end{eqnarray}
where the lapse of the $f$ metric is $n^2=\dot{f}^2+\frac{e^{-2\bar{\sigma}/M_{\rm Pl}}\alpha_\sigma \dot{\bar{\sigma}}^2}{m^2M_{\rm Pl}^2}$.
In order to be compatible with the homogeneous and isotropic Ansatz for the metric, we also assume that our matter field only depends on time $\phi=\phi(t)$. Furthermore, we introduce the following quantities for our convenience
\begin{eqnarray}\label{shortcuts}
H&\equiv& \frac{\dot{a}}{a\,N}\,, \\
r&\equiv& \frac{na}{N}\,, \\
J & \equiv &3+3(1-A)\alpha_3+(1-A)^2\alpha_4,
\end{eqnarray}
where $A$ denotes $A\equiv e^{\sigma/M_{\rm Pl}}/a$, and $H$ is the expansion rate of the physical $g$ metric, while $r$ encodes the speed of light
propagating in the $f$ metric in the units of the one propagating in the $g$ metric. The action for the extended quasidilaton (\ref{action_MG_effcoupl}) in the mini-superspace becomes
\begin{eqnarray}
\frac{S}{V} &=& \mpl^2\int dt \,a^3 N\,\Bigg\{-\Lambda-3H^2-\Lambda_A+m^2Jr(A-1)A+\frac{P_M}{\mpl^2}
+\frac{\omega \dot{\sigma}^2}{2M_{\rm Pl}^2N^2} \Bigg\}\,,
\label{eq:minisuperspace}
\end{eqnarray}
where we made use of the following definitions
\begin{equation}
\Lambda_A=m^2(A-1)\left[ J+(A-1)(\alpha_3(A-1)-3)\right]
\end{equation}
We are now in possession of all the important quantities to be to compute the background equations of motion. We do this by varying the action (\ref{eq:minisuperspace}) with respect to $N$, $a$, $\phi$, $\sigma$ and $f$. Of course as usual, the resulting system of equations of motion contains a redundant equation giving by the contracted Bianchi identity,
\begin{equation}
\frac{\partial}{\partial t} \frac{\delta S}{\delta N} - \frac{\dot{a}}{N}\frac{\delta S}{\delta a}- \frac{\dot{f}}{N}\frac{\delta S}{\delta f}- \frac{\dot{\sigma}}{N}\frac{\delta S}{\delta \sigma} - \frac{\dot{\phi}}{N}\frac{\delta S}{\delta \phi}=0\,.
\label{eq:bianchi}
\end{equation}
First of all, we start with the Friedmann equation, which can be obtained by varying the action (\ref{eq:minisuperspace}) with respect to the lapse $N$, giving rise to
\begin{equation}
3\,H^2 = \Lambda +\Lambda_A +\frac{\omega \dot{\sigma}^2}{2M_{\rm Pl}^2N^2} +\frac{\rho_M}{\mpl^2}
\label{eq:eqN}
\end{equation}
As next we vary the mini-superspace action (\ref{eq:minisuperspace}) with respect to the scale factor $a$ and combine the resulting equation with the Friedmann equation, which amounts to 
\begin{equation}
\frac{2\,\dot{H}}{N}=  \frac{(1-r)\dot{\Lambda}_A}{3H N-3\dot{\sigma}/\mpl}- \frac{\omega \dot{\sigma}^2}{M_{\rm Pl}^2N^2}-\frac{\rho_M+P_M}{\mpl^2}
\label{eq:eqa}
\end{equation}
The equation of motion for the matter field is just the standard conservation equation 
\begin{equation}
\frac{\dot{\rho}_M}{N}+3\,H(\rho_M+P_M)=0\,.
\label{eq:eqchi}
\end{equation}
Last but not least, the variation with respect to the St\"uckelberg field results in
\begin{equation}
\partial_t\left(\frac{m^2 M_{\rm Pl}^2 a^4 J(A-1)A\dot{f}}{n}\right)=0
\end{equation}
The remaining equation of motion for the $\sigma$ field reads
\begin{eqnarray}
\label{EOMsigma}
m^2\mpl N^3A(3(r-1)A(-2+\alpha_3(A-1))+J(-3+r(-1+4A))) \nonumber\\
=\omega\left( 3HN^2\dot\sigma+N\ddot\sigma-\dot{N}\dot{\sigma}\right)
\end{eqnarray}
As you can see, the dynamical background equations highly depend on the quantities $J$, $\dot{A}$ and similarly on $\dot{\sigma}$ and the dynamics of the matter field $\dot{\phi}$. In \cite{DeFelice:2013tsa}, a late-time attractor solution for the system in absence of matter fields was reported. In this late-time asymptotic solution the background quantities are either zero or constants. To be precise, $J=0$ and $H$, $A$ and $r$ are constants. With these assumptions, our background equations of course coincide with the ones in \cite{DeFelice:2013tsa}.


\section{Stability of the perturbations}\label{sec:perturbations}

The central goal of this work is to study the stability conditions for the perturbations around the full dynamical background equations from the previous section. In difference to  \cite{Motohashi:2014una}, we will assume $J\ne0$, $\dot{A}\ne0$ and $\dot{\sigma}\ne\mpl NH$, i.e. we will not assume the late-time asymptotic form of the expansion history, since we are just at the transition between matter domination and accelerating expansion making the late-time asymptotic solution not suitable to describe our current epoch. \\
Let us begin with the following perturbations for the dynamical metric $g_{\mu\nu}$
\begin{eqnarray}\label{perturbed_metric}
\delta g_{00} &=& -2\,N^2\,\Phi\,,\nonumber\\
\delta g_{0i} &=& N\,a\,\left(\partial_i B+B_i\right)\,,\nonumber\\
\delta g_{ij} &=& a^2 \left[2\,\delta_{ij}\psi +\left(\partial_i\partial_j-\frac{\delta_{ij}}{3}\partial^k\partial_k\right)E+\partial_{(i}E_{j)}+h_{ij}\right]\,.
\end{eqnarray}
where it is understood that all the metric perturbations depend on time and space. Note also that $\delta^{ij}h_{ij} = \partial^ih_{ij} = \partial^i E_i = \partial^i B_i=0$. Furthermore, we do not consider any perturbations for the St\"uckelberg fields, which fixes the gauge freedom completely. 
We perturb the matter field $\phi$ and the quasidilaton $\sigma$ as well
\begin{eqnarray}
\phi&=&\phi(t)+\mpl \delta\phi, \nonumber\\
\sigma&=& \sigma(t)+\mpl \delta\sigma \,.
\end{eqnarray}
Na\"ively counted, we encounter twelve degrees of freedom (dof) in the action \ref{action_MG_effcoupl}. Two of these dos are traceless symmetric spatial tensor fields ($h_{ij}$). In addition, we have four divergence-free spatial vector fields ($B_i$, $E_i$). The remaining six dof come in form of scalars ($\Phi$, $B$, $\psi$, $E$, $\delta\phi$, $\sigma$). In the following we will study the tensor, vector and scalar perturbations separately and establish stability conditions above the dynamical background equations.
\subsection{Tensor perturbations}
Before starting the computation of the tensor perturbations, let us first decompose the tensor field $h_{ij}$ in Fourier modes with respect to the spatial coordinates. We can perform this decomposition since our background metric is homogeneous with no spatial curvature
\begin{equation}\label{tensor_Fourier}
h_{ij}=\int \frac{\d^3k}{(2\pi)^{3/2}}h_{ij,\vec{k}}(t) \exp(i\vec{k}\cdot\vec{x}) +c.c\,.
\end{equation}
We first insert our Ansatz \ref{perturbed_metric} for the metric perturbations into the Lagrangian \ref{action_MG_effcoupl} with the fields decomposed in Fourier modes and then we use the fully dynamical background equations from the previous section. By doing that, the action quadratic in the tensor perturbations results in
\begin{equation}\label{action_tensormodes}
S^{(2)}_{\rm tensor} = \frac{\mpl^2}{8}\int d^3k\,dt\,N\,a^3\,\left[\frac{1}{N^2}\dot{h}_{ij,\vec{k}}^\star \dot{h}^{ij}_{\vec{k}}-\left(\frac{k^2}{a^2}+m_{T}^2\right)h_{ij,\vec{k}}^\star h^{ij}_{\vec{k}} \right]\,,
\end{equation}
where the mass of the tensor perturbations is given by the abbreviation
\begin{equation}
m_{T}^2\equiv\,\frac{m^2A}{(A-1)}(J(-1+rA)+A(-(2+\alpha_3)(-2+r)-(1+2\alpha_3+r)A+\alpha_3rA^2)).
\label{eq:MGW2}
\end{equation}
The tensor perturbations on top of general background solutions (\ref{action_tensormodes}) have already the right sign for the kinetic term. Similarly, they do not exhibit gradient instabilities either. Furthermore, in order to avoid tachyonic instabilities we have to impose $m_{T}^2>0$. This condition is fulfilled for the following three cases
\begin{eqnarray}\label{constraints_tensor}
0<r<A^{-1}, &\qquad& J>\bar{J} \;\; (J<\bar{J}), \qquad 0<A<1 \;\; (A>1) \nonumber\\
r>A^{-1}, &\qquad& J<\bar{J} \;\; (J>\bar{J}),  \qquad  0<A<1 \;\; (A>1)\nonumber\\
r=A^{-1}, &\qquad& A(1+\alpha_3)<2+\alpha_3, \qquad A>0
\end{eqnarray}
where $\bar{J}=\frac{A(-4+A+2(A-1)\alpha_3+r(2+A+\alpha_3-A^2\alpha_3))}{-1+rA}$. The mass term can be also expressed in terms of the dynamics of the matter field $\dot\sigma$..etc by replacing $\alpha_3$ through its expression given by the equation of motion of the dilaton $\sigma$ field \ref{EOMsigma}. 
\begin{eqnarray}
m_{T}^2&=&\frac{m^2\left(3(r-1)^2A^3-JA(-3+r+2rA)(r(2A-1)-1) \right)}{3(A-1)(r-1)} \nonumber\\
&&+\frac{\omega(rA+r-2)\mathcal{W}}{3\mpl N^3(r-1)(A-1)}
\end{eqnarray}
where we introduced the shortcut notation $\mathcal{W}=3HN^2\dot{\sigma}-\dot{N}\dot{\sigma}+N\ddot{\sigma}$. On the self-accelerating (SA) late-time asymptotic background with $J=0,\dot{A}=0, \dot{\sigma}=N\mpl H$, $\dot{H}=0$ and with the gauge choice $N=1$, the above mass term coincides with equation (38) of de Felice et al in \cite{DeFelice:2013tsa}
\begin{equation}\label{tensormass_SA}
m_{T,SA}^2= \frac{m^2(r_{SA}-1)^2A_{SA}^3+\omega H_{SA}^2(r_{SA}(A_{SA}+1)-2)}{(r_{SA}-1)(A_{SA}-1)}
\end{equation}
where $r_{SA}$ is obtained from the equation of motion for the dilaton
\begin{equation}
r_{SA}=1+\omega H_{SA}^2/(m^2 A_{SA}(\alpha_3(A_{SA}-1)-2))
\end{equation}
and $H_{SA}$ from the Friedmann equation $H^2_{SA}=\Lambda_{A_{SA}}(3-\omega/2)^{-1}$. On this background $H_{SA}$, $A_{SA}$ and $r_{SA}$ are constants. The stability of the self-accelerating background solutions requires $0<\omega <6$, and $r_{SA}>0$ for $A_{SA}>1$ or $1<r_{SA}<\bar{r}_{SA}$ for $0<A_{SA}<1$ where \cite{Kahniashvili:2014wua}
\begin{equation}
\bar{r}_{SA}  =  \frac{2}{1+A_{SA}}  
	 -  \frac{\omega}{A_{SA} (1+A_{SA})}
\left[\frac{\frac{m}{H_{SA}} (A_{SA}-1)^2}{\left(3 - \frac{\omega}{2}\right) + \frac{m^2}{H_{SA}^2} (A_{SA}-1)^2}\right]^2
\end{equation}
Furthermore, for simplicity we have assumed that the sound speed of the tensor perturbations given in \ref{action_tensormodes} is unity, i.e. $c^2_T=1$. Therefore the dispersion relation is given by\footnote{The dispersion relation in pure GR, $\mathfrak{f}_{GR}=k^2/a^2$, simply obtains an additional mass term.}
\begin{equation}
\mathfrak{f}^2 =\frac{k^2}{a^2}+m_{T}^2
\end{equation}
where $\mathfrak{f}$ is the frequency of tensor mode oscillations. Inflation generates super-horizon gravitational waves (with $\mathfrak{f}^2 \ll H^2$) that stay constant outside the horizon with an absolute value given by the amplitude of the mode at initial time (i.e. $|h_k|=A(k)$ with $A(k)=H^*/(\mpl k^{3/2})$ where $H^*$ is the Hubble expansion rate at horizon exit in a typical slow-roll inflation) . As the universe evolves (i.e. $a^2H^2$ decreases), these modes at some point re-enters the horizon (with $\mathfrak{f}^2 \sim H^2$) at time $a_{re}$ and finally end up as sub-horizon modes (with $\mathfrak{f}^2 \gg H^2$). In pure GR, the horizon crossing happens at $a_{re}^{GR}=k_{eq}^{GR}/(\sqrt{2}k)$ for momenta larger than $k_{eq}$ and $a_{re}^{GR}=(k_{eq}^{GR}/(\sqrt{2}k))^2$ for smaller momenta respectively, where $a_{eq}$ is the time of matter-radiation equality. After the horizon re-entrance the gravitational waves start oscillating with frequency $\mathfrak{f}^2 \gg H^2$ as a WKB oscillator. In pure GR, the power spectrum of sub-horizon gravitational waves today ($t=t_0$) is given by $\mathcal{P}^{GR}=2\tilde{k}^3|h^{GR}_k|^2/\pi^2$ with $\tilde{k}=a_0 \mathfrak{f}_0$. In massive gravity with the additional mass term the power spectrum becomes instead \cite{Gumrukcuoglu:2012wt}
\begin{equation}
\mathcal{P}=\frac{\mathfrak{f}_0^2}{\mathfrak{f}_0^2-m_{T,0}^2} \frac{2k^3}{\pi^2} |h_k|^2
\end{equation}
with $k=a_0\sqrt{\mathfrak{f}_0^2-m_{T,0}^2}$. Thus, the relative change of the power spectrum in massive gravity versus GR is given by\footnote{ For a detail discussion see \cite{Gumrukcuoglu:2012wt}.}
\begin{equation}\label{differencePower}
\frac{\mathcal{P}}{\mathcal{P}^{GR}}=\left(\frac{\tilde{k}/a_{\tilde k}}{k/a_k}\right)^2\frac{\mathfrak{f}_k a_k}{\mathfrak{f}_0 a_0} \frac{\mathcal{P}_{pr,k}}{\mathcal{P}_{pr,\tilde{k}}}
\end{equation}
with the primordial power spectrum $\mathcal{P}_{pr,k}=2k^3A^2/\pi^2$. In the late-time asymptotic regime, we expect that the mass term evolves into the constant value $m_T \to m_{T,SA}$ given by equation (\ref{tensormass_SA}). However, we need the full solution of the background equations in order to see how $m_T$ evolves in time and to be able to compute its value today and also during matter domination. It would be very interesting to solve \ref{differencePower} numerically and study the impact of the additional dilation field (for the case in the pure standard massive gravity see \cite{Gumrukcuoglu:2012wt}). In the standard pure massive gravity theory the mass of the graviton is constrained to be of the same order as the Hubble expansion rate today $m\sim H_{0}$. Therefore this severely restricts $m_{T}$ to be of the same order $m_{T}\sim H_{0}$. In the quasi dilation case this can be avoided. Even if  $m\sim H_{0}$, one can have $m_{T}\gg H_{0}$ due to the presence of the additional degree of freedom. A natural outcome from this might be the large-angle suppression of power in the microwave background \cite{Kahniashvili:2014wua}. This together with numerical solutions and comparison to current data will be studied elsewhere.

\subsection{Vector perturbations}
We would like now pay our attention to the stability conditions of the vector perturbations. As we did for the tensor perturbation, we first 
decompose the vector modes $E_i$ and $B_i$ in Fourier modes 
\begin{eqnarray}
E_i=\int \frac{d^3k}{(2\pi)^{3/2}} E_{i,\vec{k}}\;(t)e^{i\vec{k}\cdot\vec{x}} +c.c, \;\;\;\;\;\;\;\;\;\;\;     B_i=\int \frac{d^3k}{(2\pi)^{3/2}} B_{i,\vec{k}}\;(t)e^{i\vec{k}\cdot\vec{x}} +c.c.
\label{fourierEandB}
\end{eqnarray}
As next we expand the Lagrangian \ref{action_MG_effcoupl} to second order in the vector perturbations. Note that not all of the vector fields are dynamical, indeed the vector fields $B_i$ do not have any kinetic terms. We can therefore compute the equation of motion with respect to $B_i$ and integrate them out
\begin{equation}
B_{i,\vec{k}}=\frac{k^2a(1+r)}{2N(k^2+k^2r+2m^2a^2A(J-(2+\alpha_3)A+\alpha_3A^2))}\dot{E}_{i,\vec{k}}.
\end{equation}
After using this expression, the quadratic action in the vector perturbations becomes
\begin{equation}\label{action_vectormodes}
S^{(2)}_{\rm vector} = \frac{\mpl^2}{16}\int d^3k\,dt\,k^2a^3N \left[m_V^2
\dot{E}_{i,\vec{k}}^\star\dot{E}^i_{\vec{k}}-m_{T}^2 E_{i,\vec{k}}^\star E^i_{\vec{k}} 
\right]\,.
\end{equation}
where we introduced for convenience the shortcut notation
\begin{equation}
m_V^2 \equiv \frac{2m^2a^2A(J+A(-2+\alpha_3(A-1)))}{N^2(k^2+k^2r+2m^2a^2A(J-(2+\alpha_3)A+\alpha_3A^2))}
\end{equation}
For the stability of vector perturbations on top of a generic background solution we have to impose the right sign for the kinetic and gradient terms. For the absence of ghost instability, we require that the kinetic term has the right sign. This is the case if we impose $m_V^2>0$. This on the other hand requires $r>0$, $A>0$ and $J>A(2+\alpha_3(1-A))$ or $k^2(1+r)+2a^2m^2A(J+A(-2+(A-1)\alpha_3))<0$. Similarly as for the tensor perturbations the absence of gradient and tachyonic instability requires $m_T^2>0$, which are the same constraints as in \ref{constraints_tensor}.\\
Again we can replace the dependence on $\alpha_3$ through the other parameters of the theory using the equation of motion for the dilaton field. In this case the prefactor in front of the kinetic term becomes
\begin{equation}
m_V^2=\frac{1}{N^2}-\frac{3k^2\mpl N(r^2-1)}{\mpl N^3(-3k^2+r(3k^2r-8m^2a^2J(A-1)A))+2\omega a^2 \mathcal{W}}
\end{equation}
Recall that $\mathcal{W}=3HN^2\dot{\sigma}-\dot{N}\dot{\sigma}+N\ddot{\sigma}$. On the self-accelerating background with $J=0,\dot{A}=0, \dot{\sigma}=N\mpl H$, $\dot{H}=0$ and with the gauge choice $N=1$, the expression for $m_V^2$ becomes simply
\begin{equation}
m_{V,SA}^2=\frac{2\omega a^2 H_{SA}^2}{N^2(2\omega a^2 H_{SA}^2+k^2(r_{SA}^2-1))}
\end{equation}
which coincides with the expression of de Felice et al  \cite{DeFelice:2013tsa}. 

\subsection{Scalar perturbations}
Now we will be concentrating on the stability of the scalar perturbations in the extended quasi-dilaton massive gravity model with matter field. As we mentioned above, six degrees of freedom appear in form of scalar fields ($\psi$, $\delta\sigma$, $\delta\phi$, $E$, $B$, $\Phi$ ). We first expand the action \ref{action_MG_effcoupl} to quadratic order in the scalar perturbations and introduce their Fourier modes
\begin{eqnarray}
\Phi=\int \frac{d^3k}{(2\pi)^{3/2}} \Phi_{\vec{k}}\;(t)e^{i\vec{k}\cdot\vec{x}} +c.c, && \;\;\;\;\;\;\;\;\;\;\;     B=\int \frac{d^3k}{(2\pi)^{3/2}} B_{\vec{k}}\;(t)e^{i\vec{k}\cdot\vec{x}} +c.c. \nonumber\\
\psi=\int \frac{d^3k}{(2\pi)^{3/2}} \psi_{\vec{k}}\;(t)e^{i\vec{k}\cdot\vec{x}} +c.c, &&\;\;\;\;\;\;\;\;\;\;\;     E=\int \frac{d^3k}{(2\pi)^{3/2}} E_{\vec{k}}\;(t)e^{i\vec{k}\cdot\vec{x}} +c.c.  \nonumber\\
\delta\phi=\int \frac{d^3k}{(2\pi)^{3/2}} \delta\phi_{\vec{k}}\;(t)e^{i\vec{k}\cdot\vec{x}} +c.c, &&\;\;\;\;\;\;\;\;\;\;\;   \delta\sigma=\int \frac{d^3k}{(2\pi)^{3/2}} \delta\sigma_{\vec{k}}\;(t)e^{i\vec{k}\cdot\vec{x}} +c.c
\label{fourierPhiandpsi}
\end{eqnarray}
The corresponding kinetic matrix (and the Hessian matrix) contains two vanishing eigenvalues, signalling the existence of two constraint equations which make two out of the six scalar fields not propagating
\begin{eqnarray}
\mathcal{K}_{\psi, \delta\sigma, \delta\phi, E, B, \Phi}=
\begin{pmatrix}
-6&0&0&0 &0&0\\
0&\omega &0&0 &0&0\\
0&0&2P(X_\phi)' +4P(X_\phi)''X_\phi &0 &0&0\\
0&0&0&k^4/6&0&0 \\
0&0&0&0&0&0 \\
0&0&0&0&0&0
 \end{pmatrix}
\end{eqnarray}
Since the quadratic action does not have any kinetic term for the scalar fields $\Phi$ and $B$, we can compute their equations of motion in order to obtain the corresponding two constraint equations
\begin{eqnarray}
\frac{3\mpl aN QA}{(A-1)^2(r+1)}B_{\vec{k}}+6\mpl HN \Phi_{\vec{k}}-k^2\mpl \dot{E}_{\vec{k}}-6\mpl  \dot{\psi}_{\vec{k}} =
3\omega \dot{\sigma}\delta\sigma_{\vec{k}}+6P(X_\phi)'\dot{\phi}\delta\phi_{\vec{k}} 
\end{eqnarray} 
and  similarly 
\begin{eqnarray}
&&k^4\mpl^2N E_{\vec{k}}+3(2k^2\mpl^2 aB_{\vec{k}}HN+2k^2\mpl^2N\psi_{\vec{k}}+a^2(-\frac{3\mpl^2NQA\delta\sigma_{\vec{k}}}{(A-1)^2} -6\mpl^2H^2N\Phi_{\vec{k}} \nonumber\\
&&+6\mpl^2H\dot{\psi}_{\vec{k}}+\frac{1}{N^3}(\frac{3\mpl^2N^4AQ\psi_{\vec{k}}}{(A-1)^2}+N^2(-\mpl\omega\dot{\sigma}\dot{\delta\sigma}_{\vec{k}}+\omega\Phi_{\vec{k}}\dot{\sigma}^2-2\mpl\dot{\phi}P(X_\phi)'\dot{\delta\phi}_{\vec{k}} \nonumber\\
&&+2\Phi_{\vec{k}}\dot{\phi}^2P(X_\phi)')+4\dot{\phi}^3(-\mpl \dot{\delta\phi}_{\vec{k}}+\Phi_{\vec{k}}\dot{\phi})P(X_\phi)'')))=0
\end{eqnarray} 
We can solve them for $B$ and $\Phi$. For clarity, we have defined the quantity
 \begin{equation}
 Q=-m^2J(-1+A)+(\Lambda_A+m^2(A-1)^2)A.
 \end{equation}
 After plugging back the solutions for $B$ and $\Phi$, the resulting action depends only on the remaining four scalar fields ($\psi$, $\delta\sigma$, $\delta\phi$, $E$). The kinetic matrix of these four remaining scalar fields still has a vanishing determinant, meaning that there is still one more constraint that can be used to integrate out one of the scalar fields. In fact, the sub-kinematic matrix for the fields ($\psi$, $\delta\sigma$, $\delta\phi$) has already a vanishing determinant and it has one zero eigenvalue $\lambda_1=0$ and two non-vanishing eigenvalues $\lambda_2$ and $\lambda_3$
\begin{eqnarray}\label{3newvarMatrix}
\mathcal{K}_{\psi, \delta\sigma, \delta\phi}=
\begin{pmatrix}
0&0&0\\
0&\lambda_2 &0\\
0&0&\lambda_3
 \end{pmatrix}
\end{eqnarray}
The corresponding eigenvectors to the three eigenvalues $\lambda_1$, $\lambda_2$ and $\lambda_3$ are respectively given as
\begin{eqnarray}
v_{1}=
\begin{pmatrix}
\mpl H N/\dot{\phi}\\
\dot{\sigma}/\dot{\phi}\\
1
 \end{pmatrix}, \qquad  v_{2}=
\begin{pmatrix}
v_{21}\\
v_{22}\\
1
 \end{pmatrix},  \qquad  v_{3}=
\begin{pmatrix}
v_{31}\\
v_{32}\\
1
 \end{pmatrix}
\end{eqnarray}
We omit the cumbersome expressions for the eigenvalues and eigenvectors here. In the UV limit for instance we manage to write down the eigenvalues in a manageable way 
\begin{equation}
\lambda^{UV}_1=0, \quad \lambda^{UV}_{2,3}=\frac{1}{2\mpl^2H^2N^4}\left(\frac{p}{2}\pm \sqrt{\left(\frac{p}{2}\right)^2-q}\right)
\end{equation}
where
\begin{eqnarray}
\frac{p}2&=&\mpl^2\omega H^2N^4+2\mpl^2H^2N^4P(X_\phi)'+N^2\omega\dot{\sigma}^2+2N^2\dot{\phi}^2P(X_\phi)' \nonumber\\
&&+4\mpl^2H^2N^2\dot{\phi}^2P(X_\phi)''+4\dot{\phi}^4P(X_\phi)'' \nonumber\\
q&=& 8\mpl^2\omega H^2N^4(\mpl^2H^2N^2+\dot{\sigma}^2+\dot{\phi}^2)(N^2P(X_\phi)' +2\dot{\phi}^2P(X_\phi)'')
\end{eqnarray}
Before moving on, at this stage it is useful to remind the reader that these perturbations are performed around the general background equations including the matter fields. Comparing our analysis with \cite{Motohashi:2014una}, it is clear that the difference comes fro the fact that the authors in  \cite{Motohashi:2014una} did not perform their analysis using the full dynamical background equations. The dynamical background evolution dictates that $\dot{\sigma}=\mpl\left( HN+\frac{\dot{A}}{A}\right)$ and not $\dot{\sigma}=\mpl HN$. Dynamical terms of the form $\dot{A}/A$ are missing in  \cite{Motohashi:2014una}. At this stage, it is not clear to us how to justify to use the late-time asymptotic form of the expansion history to evaluate the perturbations for our current epoch. The full dynamical background equations would need to be solved numerically and the perturbations should be performed about these exact solutions. \\

The presence of the zero eigenvalue in \ref{3newvarMatrix} indicates that we can find a constraint equation to eliminate one of the degrees of freedom,  the would-be Boulware-Deser ghost. After integrating it out, the resulting action can be expressed in the following form
\begin{equation}
S^{(2)}_{\rm scalar}= \frac{\mpl^2}{2}\int d^3k \,dt\,a^3 \,\left(\dot{\Pi}^\dagger\,\hat{K}\,\dot{\Pi} + \dot{\Pi}^\dagger\,\hat{{\cal N}}\,\Pi- \Pi^\dagger\,\hat{{\cal N}}\,\dot{\Pi}-\Pi^\dagger\,\hat{M} \,\Pi\right)\,,
\end{equation}
where $\Pi$ denotes the three physical propagating degrees of freedom $\Pi=\{ \pi_{1,\vec{k}}, \pi_{2,\vec{k}},\pi_{3,\vec{k}}\}$. The expressions for the kinetic, gradient matrices ..etc are very cumbersome around general background solutions after integrating out the would-be Boulware-Deser ghost, even in their UV and IR limits, which makes it very hard to write them down in a handy way. However, the for us important conditions here are the ones coming from the ghost absence. For this, we have to guarantee that the kinetic matrix is positive definite. This on the other hand imposes conditions on the determinant of the matrix $\det{K}>0$ and parts of it $\det{K_{33}}>0$, $K_{22}K_{33}-K_{23}^2>0$. By requiring these conditions on the kinetic matrix in the UV and IR limit together amounts to\footnote{In the self accelerating late-time asymptotic solution the condition is the same as reported in \cite{DeFelice:2013tsa}.}
\begin{equation}
0<\omega < 6  \qquad  {\rm and}  \qquad A^2<\frac{\alpha_\sigma}{m^2} \left(H+\frac{\dot{A}}{A}\right)^2 < r^2A^2
\end{equation}
As you can see, this condition differs from the one derived in \cite{Motohashi:2014una}. Since they were assuming the late-time asymptotic form of the expansion history, terms proportional to $J$, $\dot{A}$ and $\dot{\sigma}$ were missing in their analysis. However, one needs to work with the full solution of the dynamical background equations to describe the evolution of the universe. The reported instability in \cite{Motohashi:2014una} was probably just due to the missing terms $\dot{A}/A$ based on the late-time approximated solution. It would be very interesting to study numerical solutions of the general background equations and check the consequences of the stability conditions on the parameters. The exact numerical solution and comparison to the current data on cosmic expansion is being studied elsewhere.

\section{Conclusions}
In this work, we have considered the extended quasidilaton massive gravity theory, which includes an additional scalar degree of freedom on top of the massive graviton. This extended version of the quasidilaton has an additional parameter, which was introduced to give rise to stable self-accelerating solution at late times. On top of these self-accelerating background it was shown that the perturbations of the tensor, vector and scalar perturbations were stable after imposing the necessary conditions on the parameters of the theory \cite{DeFelice:2013tsa}. Here we extend their stability analysis to general background including the presence of matter fields. \\

After discussing the general dynamical background equations, we studied the stability conditions on the tensor, vector and scalar perturbations. The tensor perturbations have automatically the right sign for the kinetic and gradient term without imposing further constraints. The only constraint coming from the tensor perturbations is the positivity of the mass term $m_T^2>0$. From this condition we were able to put constraints on the background quantities $J$, $r$, ...etc and on the free parameters of the theory. From the vector perturbations we obtain an additional constraint coming from the coefficient in front of the kinetic term, $m_V^2>0$ in order to avoid ghost instability. This condition puts further constraint on $J$ and the parameter $\alpha_3$ together with the condition $r>0$. The absence of gradient instability of the vector perturbations does not give any further constraint since it is the same constraint as for the absence of tachyonic instability of tensor perturbations. Last but not least, the stability of the scalar perturbations not only imposes constraints on the parameter $\omega$ and $\alpha_\sigma$ but also on the dynamical quantities $\dot{A}$. Again for the stability of the scalar perturbations, the parameter $\alpha_\sigma$ plays a crucial role. An important step in future works would be to study the dynamical system of cosmological solutions by using phase map analysis and explore the critical points of the cosmological equations as well as numerical solutions.

\label{sec:conclusion}

\acknowledgments

We would like to thank Tomi Koivisto, Shinji Mukohyama and Norihiro Tanahashi for very useful discussions.

\appendix

	\bibliographystyle{JHEPmodplain}
	\bibliography{EffectiveCoupling_cosmology}

\end{document}